\documentclass[11pt]{article}

\usepackage[utf8]{inputenc}
\usepackage[english]{babel}
\usepackage{color}

\usepackage{xr}
\usepackage{soul}

\usepackage{CJKutf8}
\usepackage{titling}

\usepackage{xpatch}
\makeatletter
\newenvironment{mytitlepage}%
  {\begin{titlepage}\def\@thanks{}}%
  {\end{titlepage}}
\xpatchcmd\titlepage{\setcounter{page}\@ne}{}{}{}
\xpatchcmd\endtitlepage{\setcounter{page}\@ne}{}{}{}
\makeatother

\usepackage{amsfonts,amsmath,amssymb,amsbsy}
\usepackage{mathpazo}
\usepackage{mathptmx}

\usepackage{graphicx}
\usepackage{pdflscape}

\usepackage{tabularx}
\usepackage{dcolumn} 
\usepackage{booktabs}
\usepackage{threeparttable}
\usepackage{comment}

\usepackage[justification=justified,singlelinecheck=false]{caption}
\usepackage{subcaption}
\usepackage{float}
\restylefloat{table}
\restylefloat{figure}

\usepackage{xcolor}
\usepackage{footnotebackref}
\usepackage{hyperref}
\hypersetup{
    colorlinks=true,
    citecolor=black,
    linkcolor=black,
    urlcolor=black
}
\usepackage{enumitem}

\usepackage[square,numbers]{natbib}

\makeatletter
\newcommand*{\addFileDependency}[1]{
  \typeout{(#1)}
  \@addtofilelist{#1}
  \IfFileExists{#1}{}{\typeout{No file #1.}}
}
\makeatother

\newcommand*{\myexternaldocument}[1]{%
    \externaldocument{#1}%
    \addFileDependency{#1.tex}%
    \addFileDependency{#1.aux}%
}

\myexternaldocument{si_pnas} 

\usepackage{setspace}

\usepackage{geometry}
\usepackage{afterpage}

\evensidemargin\oddsidemargin
\begin{document}

\begin{mytitlepage}
\title{Quantifying the Influence of Climate on Human Mind and Culture: Evidence from Visual Art}\author{Shuhei Kitamura\thanks{%
Institute of Social and Economic Research, Osaka University, 6-1 Mihogaoka, Ibaraki, Osaka 567-0047, Japan. Email: kitamura@iser.osaka-u.ac.jp.}}
\date{}
\maketitle

\begin{abstract}
While connections between climate and the human mind and culture are widely acknowledged, they are not thoroughly quantified. Analyzing 100,000 paintings and data on 2,000 artists from the 13th to 21st centuries, the study reveals that the lightness of the paintings exhibited an interesting U-shaped pattern mirroring global temperature trends. There is a significant association between the two, even after controlling for various factors. Event study analysis using the artist-level data further reveals that high-temperature shocks resulted in brighter paintings in later periods for artists who experienced them compared to the control group. The effect is particularly pronounced in art genres that rely on artists' imaginations, indicating a notable influence on artists' minds. These findings underscore the enduring impact of climate on the human mind and culture throughout history and highlight art as a valuable tool for understanding people's minds and cultures. 
\end{abstract}
\end{mytitlepage}

\clearpage


\section{Introduction\label{sec:intro}}

To understand significant social changes like political transitions and economic development, comprehending the role of people's minds and cultures is essential. This includes exploring the impact of emotions during specific social changes, such as the overthrow of reigns, and identifying triggering factors for psychological responses. Previous studies have discussed the effects of climate on political transitions and the demise of ancient empires and dynasties (e.g., the Akkadian Empire \cite{cullen2000}, Classic Maya civilization \cite{haug2003, kennett2012}, Chinese dynasties \cite{zhang2006, zhang2014}, Angkor \cite{buckley2010}, the Western Roman Empire \cite{butgen2011}, and contemporary sub-Saharan Africa \cite{bruckner2011}), suggesting that social instability caused by climate-driven economic difficulties is a plausible driving force (see \cite{hsiang2013} for a review). Climatic shocks may also influence people's psychology, evoking emotions such as anger and fear. Findings from previous studies indicate a direct link between climate and ``aggressive" behaviors such as horn honking \cite{kenrick1986}, sports violence \cite{larrick2010}, and swearing on social media \cite{baylist2018}. Laboratory experiments have also shown that high temperatures can lead to heightened destructive behavior \cite{almas2020}. In addition, an indirect link may exist where emotional responses arise owing to inadequate responses from ruling authorities during climate change crises. Quantitative measures are needed to assess people's minds and cultures in historical contexts.

Previous studies have used text data to measure people's minds and cultures. Analyzing the language in millions of books shows that words associated with emotion have declined in the last century \cite{acerbi2013, morin2017, scheffer2021}. The bias that people use more positive than negative words has also declined \cite{iliev2016}. Other studies using text data have found changes in morality \cite{kesebir2012}, individualism \cite{greenfield2013, grossmann2014}, gender and ethnic stereotypes \cite{garg2018}, subjective well-being \cite{hills2019}, prosociality \cite{martins2020}, and cultural tightness-looseness \cite{jackson2019, choi2022} (see also \cite{atari2023} for a review). While text data are valuable sources for measuring people's minds and cultures, studies often focus on relatively recent periods or selected countries, possibly owing to current data availability. Furthermore, text data may not fully capture some aspects of the human mind and culture.

To address these challenges, this study proposes using art, particularly paintings, as an alternative tool for measuring people's minds and cultures, complementing existing text data. The idea of using art to infer people's minds and cultures dates back to Georg Wilhelm Friedrich Hegel, who used art to infer the {\it zeitgeists} of its time of creation \cite{hegel1998}. According to Tolstoy, art has a unique ability to convey artists' feelings \cite{tolstoy1930, winner1982}. Art historian and sociologist Arnold Hauser attempted to infer cultural values and norms in society from an analysis of artworks \cite{hauser1999}.

This study compiled digital images of 100,000 paintings from the late 13th century to the early 21st century and biographical data of over 2,000 artists. Combining artist’s biographical data with their artworks enabled us to assign information on when and where each artwork was created. This allowed for the measurement of the characteristics of the artwork in certain periods and geographical units (e.g., the average brightness of paintings in the 19th century in France). The present study used this unique dataset to analyze the influence of climate on the human mind and culture.

Although art can potentially be used to measure the human mind and culture in various ways, this study focused on analyzing its color, especially brightness, as it is a fundamental element of art associated with emotional valence \cite{valdez1994, wilms2018, jonauskaite2019}. The first analysis investigated how the lightness of paintings changed over time, revealing an interesting U-shape corresponding to global temperature patterns. The pattern roughly corresponds to the pattern of each period, approximately coinciding with the Medieval Warm Period, the Little Ice Age, and the period of recent global warming. Using the data at the country level, a statistically significant association between these factors was found, even after controlling for various factors. The significance remained after restricting the sample to the periods before the Industrial Revolution, indicating that industrialization or recent hot temperatures are not the main drivers of the result.

While the country-level analysis revealed a significant association between temperature anomalies and the lightness of paintings, it might obscure interesting variations among individual artists. In addition, the results may still be subject to endogeneity issues. Therefore, the second set of analyses applied econometric techniques to the data at the artist level to address these concerns. The first analysis applied an event study analysis with artist-, country-, and year-fixed effects, along with other controls. After controlling for such factors, this analysis examined what would happen to the lightness of paintings if artists were unintentionally exposed to historically hot temperatures during their lifetimes.

The analysis revealed that while there were no systematic changes in the lightness of paintings before the high-temperature shocks, the paintings became {\it brighter} after the artists had been exposed to these shocks, compared to the control group. This effect was particularly pronounced in paintings that strongly rely on the artists’ imaginations, such as religious and mythological paintings. In contrast, no similar effects were found for paintings that rely more on real objects, such as landscapes and portraits, indicating that the influence of temperature shocks comes through people’s minds, such as emotions, sentiments, and mood, rather than simply reflecting the outside atmosphere. In addition, the study provided supporting evidence that the paintings became {\it darker} after cold-temperature shocks.

Finally, the study also examined heterogeneous responses based on the characteristics of artists using the difference-in-differences method. The analysis found that self-trained artists are more likely to respond to temperature shocks, while academically trained artists are less likely. However, no significant heterogeneous responses were found with respect to the artists’ age, sex, or income sources.

\section{Results\label{sec:results}}

\subsection{Analysis at the Country Level}

The top panel of Figure \ref{fig:main}A shows the lightness of the painting (HSV-V) between 1270-2006, revealing an interesting U-shaped pattern: the paintings were initially brighter, then darker, and eventually brighter again. The middle panel in Figure \ref{fig:main}A displays the mean surface temperature anomalies, defined relative to the 1961-1990 reference period mean. A period of coldest temperatures is known as the Little Ice Age, describing the period between the 14th and 19th centuries \cite{white2013}. These global patterns are also similar across different countries (Figure \ref{fig:main}B), which is somewhat surprising considering the spatial segregation of art materials and methods, especially in earlier periods before the advancement of transportation and communication technology.

The top-left panel of Figure \ref{fig:main}C shows a partial correlation plot after removing the influences of economic development, population changes, political disturbances, country-specific characteristics that remain relatively constant over time, such as geography and nationality, and art movements by country. Even after controlling for such factors, the relationship remains highly statistically significant ($r$ = 0.048, $p$ $<$ 0.001). To assess whether this is driven by the upper tail of the distribution of temperature anomalies in recent decades, the panel below shows the same partial correlation plot but with the sample restricted to the pre-Industrial Revolution (IR) periods (before 1750). Even in this restricted sample, the relationship remains highly significant ($r$ = 0.138, $p$ $<$ 0.001), indicating that industrialization and recent hot temperatures are not the main drivers of the result. Conversely, when restricting to the post-IR periods, similar or slightly weaker results are obtained ($r$ = 0.033, $p$ = 0.006). The results without the control variables are also presented in the Supplementary Material ({\it SI} Appendix, Figure \ref{fig:corr_no_control}).

The right panels of Figure \ref{fig:main}C mirror the respective left panels by categorizing the (residualized) temperature anomalies into high and low groups based on the median value. In both the full and restricted samples, paintings created in the high-temperature years tend to be brighter than those made in the low-temperature years, even after removing the influences of the factors mentioned above.

However, there are several concerns regarding the above results. One concern is that paintings might have become darker simply because painting styles have changed, although why painting styles themselves have evolved in the manner observed today remains an interesting question. An example used during the dark period of paintings is {\it chiaroscuro}, a technique that was developed during the Renaissance period and involves a dramatic contrast between light and shadow. A related but more dramatic form of chiaroscuro is {\it tenebrism}, which was introduced and developed during the Baroque period. Artists like Caravaggio (1571-1610), Georges de La Tour (1593-1652), and Rembrandt (1606-1669) are known for using this style \cite{chilvers2009}. To check whether the above phenomenon is driven by different painting styles, examples of paintings from each period, which are colored gray in Figure \ref{fig:main}A (1400-1499, 1600-1699, and 1800-1899), are presented in Figure \ref{fig:main}D. In this figure, the paintings at the top are brighter paintings (90th percentile), those in the middle are paintings with average lightness, and those at the bottom are darker paintings (10th percentile). Consistent with this explanation, one of the darkest paintings in the 17th century, Georges de la Tour's {\it The Adoration of the Shepherds} (c.1644), employs this style. This suggests that the darkness in paintings may be partially attributed to a change in painting style, especially at the bottom tail. However, the figure also shows that the upper tails of paintings, not necessarily using this style, tend to be darker. Furthermore, as demonstrated earlier, controlling for art movements by country still produces significant results (Figure \ref{fig:main}C). These results indicate that changes in painting styles may contribute to explaining the observed relationship to some extent, but they do not fully account for it.

Similarly, the number of paintings in the dataset does not explain this pattern (bottom panel in Figure \ref{fig:main}A). In addition, alternative definitions of lightness show very similar patterns ({\it SI} Appendix, Figure \ref{fig:lightness_alt}A), whereas the other elements of color in the HSV color space, i.e., saturation and hue, do not exhibit similar patterns ({\it SI} Appendix, Figure \ref{fig:lightness_alt}B).

\subsection{Analysis at the Artist Level}
While the analysis above revealed a significant association between climate and the lightness of paintings, there are still some concerns. First, the results may still be subject to endogeneity issues, such as unaccounted confounders. Second, the aggregated data might obscure significant variations among artists. Therefore, the subsequent analyses employ an econometric approach using artist-level data to address these concerns.

\subsubsection{Influence of temperature shocks on artists}\label{sec:event_study}
The first analysis employs an event study analysis with artist-, country-, and year-fixed effects using a panel dataset of artists. These fixed effects account for common factors within an artist (e.g., personality and educational background), within a country (e.g., geography and nationality), and within a year (e.g., the availability of particular art materials and methods, common macroeconomic shocks, and political shocks), respectively. Controlling for such potential confounders, the regression analysis examines what would happen to the lightness of paintings if artists had been unintentionally exposed to historically hot temperatures during their lifetimes. The study focuses on hot-temperature shocks because cold-temperature shocks in early periods rarely align with artworks in the dataset, posing challenges for a rigorous analysis. The results of cold temperature shocks are presented and discussed in the Supplementary Material ({\it SI} Appendix, Section \ref{sec:cold}).

Figure \ref{fig:event_main}A displays the coefficients from the regression and the associated 95\% confidence intervals, where zero on the x-axis represents the year when artists were first exposed to historically hot temperatures in their lifetimes. Shocks were defined as the top 1\% of global temperature anomalies ({\it SI Appendix}, Section \ref{sec:methods}). The figure illustrates that before experiencing the temperature shock, the lightness of the paintings was very similar between the treatment and control groups. After the shocks, however, the treated artists (those who experienced hot temperature shocks for the first time in their lives) tended to create brighter paintings compared to the control group. The effect is clear immediately after the shock and becomes statistically significant after several years. The magnitude of the effect was 0.28 standard deviations ($p$ $<$ 0.001) after 20 years from the shock.

The Supplementary Material includes regression results that additionally control for the covariates of economic development, population changes, and political disturbances ({\it SI Appendix}, top-left panel in Figure \ref{fig:event_study_robust}). Furthermore, the Supplementary Material includes regression results that take into account correlations in the error term ({\it SI Appendix}, top-right panel in Figure \ref{fig:event_study_robust}), the results using a different estimation method \cite{sun2021} ({\it SI Appendix}, middle-left panel in Figure \ref{fig:event_study_robust}), the results using a subset of the sample ({\it SI Appendix}, middle-right panel in Figure \ref{fig:event_study_robust}. See also Section \ref{sec:dist_shocks} in the Supplementary Material), and the results after removing very bright pixels ({\it SI Appendix}, bottom-left figure in Figure \ref{fig:event_study_robust}. See also Section \ref{sec:methods} in the Supplementary Material). Overall, the analyses yield very similar results.

Finally, the Supplementary Material also includes an analysis of artists' ``productivity,” measured by the number of paintings for each artist and year in the dataset ({\it SI Appendix}, Section \ref{sec:productivity}). The results provide no clear evidence that temperature shocks affect productivity.

\subsubsection{Heterogeneous responses across artists}

Artists may respond differently to temperature shocks depending on their age, sex, source of income, and how they learned to paint. The following analysis examines these heterogeneous responses using econometric techniques. Specifically, for each personal attribute, e.g., age, the lightness of paintings is regressed on the same temperature shocks interacting with the attribute, along with other variables, using a difference-in-differences (DID) estimation ({\it SI Appendix}, Section \ref{sec:methods}).

\begin{figure}
\begin{center}
\includegraphics[width=0.8\linewidth]{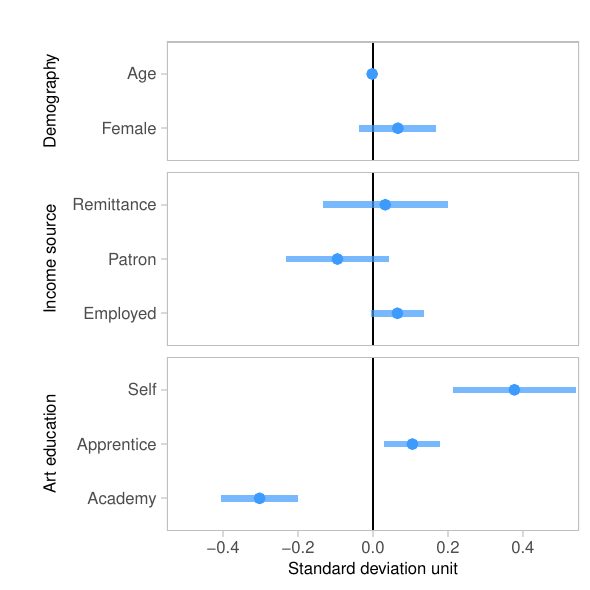}
\caption{Effects of personal attributes. The estimated coefficients and associated 95\% CIs of the interaction term ($D_{art} \times Z_{a}$) in equation (\ref{eq:event_hetero}).\label{fig:attribute}}
\end{center}
\end{figure}

Figure \ref{fig:attribute} shows the estimated coefficients and associated 95\% confidence intervals of the interaction terms for each attribute. The figure reveals that the effect of hot temperature shocks does not systematically differ by artists’ age ($b$ = -0.002, $p$ = 0.084), gender ($b$ = 0.066, $p$ = 0.206), or income source (remittances: $b$ = 0.033, $p$ = 0.702, patrons: $b$ = -0.095, $p$ = 0.173, employed: $b$ = 0.065, $p$ = 0.067). In contrast, there are differences depending on how artists learned to paint: self-taught artists ($b$ = 0.377, $p$ $<$ 0.001) and those who learned to paint through apprenticeship ($b$ = 0.105, $p$ = 0.006) are more likely to respond to shocks, whereas those who learned to paint at an academy ($b$ = -0.303, $p$ $<$ 0.001) are less likely to be affected. One plausible interpretation is that academy-trained artists are influenced by what has been taught at an academy and find it difficult to deviate from it. In contrast, other types of artists, particularly those who learned to paint on their own, could relatively freely adjust their painting styles when affected by temperature shocks.

\subsection{Mechanism}
To further understand the underlying mechanism behind the above findings, the last analysis examines whether changes in the lightness of paintings occur through either the internal channel, i.e., the effect through artists’ minds, or the external (or mechanical) channel, i.e., artists simply depicting bright scenes outdoors, brightly dressed figures, or both. For example, Dutch paintings from the Little Ice Age often depict scenes from winter landscapes \cite{behringer2010, degroot2018}. The following analysis uses particular art genres to narrow down the possible channels.

In paintings, some references are made to things that exist in reality, whereas others depend on the artist’s imagination. Therefore, if the effect appears in art genres that rely more on real objects, such as landscapes and portraits, it is more likely that artists depicted what they saw, indicating an external channel. However, if the effect appears in art genres that rely more on the artist’s imagination, such as religious and mythological paintings, it is more likely to suggest an internal channel.

\begin{figure}
\begin{center}
\includegraphics[width=\linewidth]{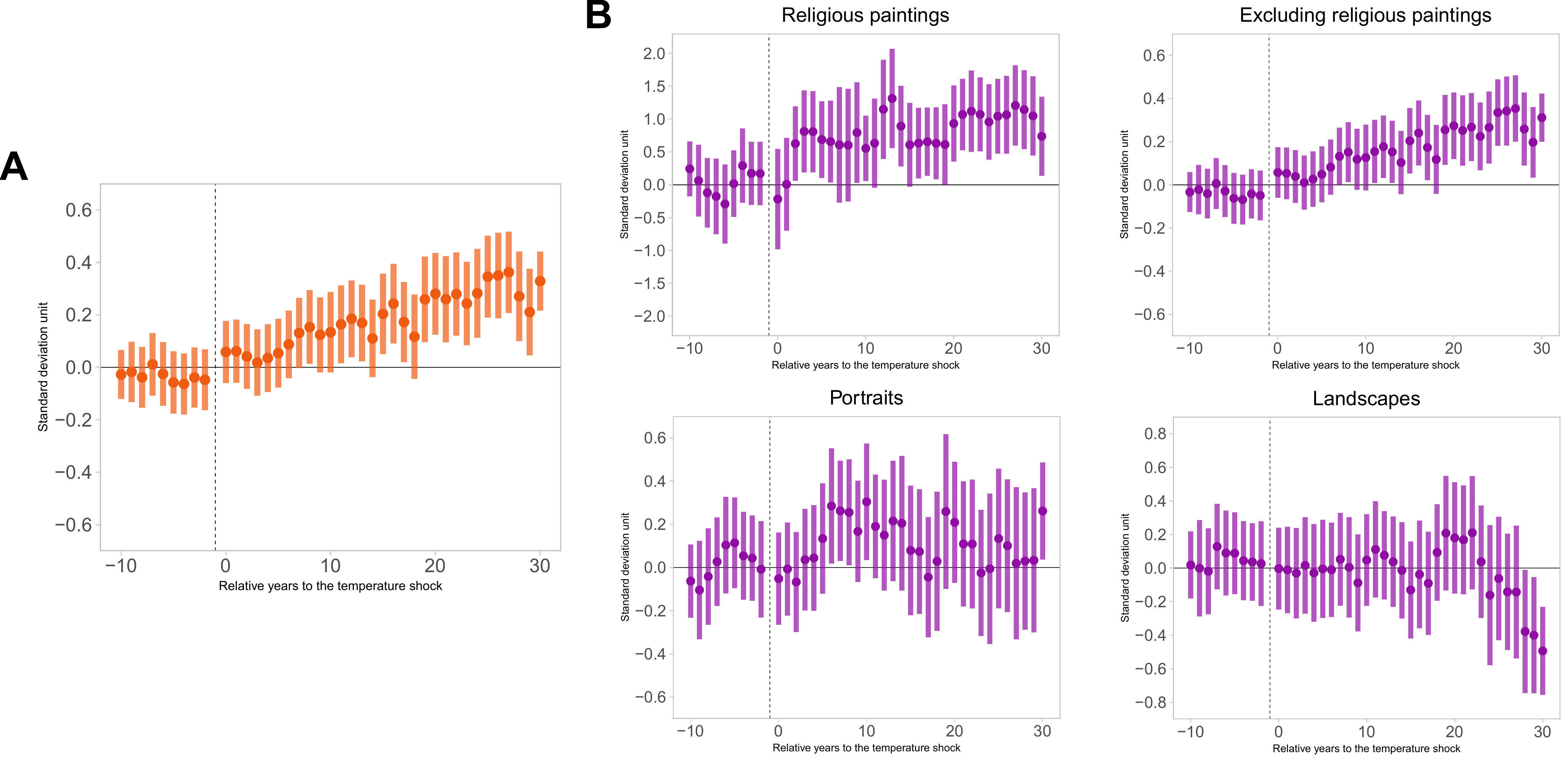}
\caption{Effects of hot-temperature shock on painting lightness. (A) Estimated coefficients and associated 95\% CIs using equation (\ref{eq:event_main}). (B) Same as (A), but the paintings in the sample are limited to certain genres: religious and mythological paintings (top left), all genres except religious or mythological paintings (top right), portraits (bottom left), and landscapes (bottom right).\label{fig:event_main}}
\end{center}
\end{figure}

The top-left panel of Figure \ref{fig:event_main}B utilizes the same regression equation as that used in Figure \ref{fig:event_main}A in Section \ref{sec:event_study}, with the distinction that the data consist solely of religious and mythological paintings. Similar to the previous result, the figure shows that artists tended to draw brighter paintings after hot temperature shocks. They responded relatively immediately, and the effect size was greater than that shown earlier (0.93 standard deviations ($p$ = 0.002) after 20 years from the shocks). However, this does not imply that previous results were driven only by these particular art genres. The top-right panel in Figure \ref{fig:event_main}B shows that excluding these paintings does not significantly change the main findings. In contrast, no clear pattern was observed for either portraits (bottom-left panel in Figure \ref{fig:event_main}B) or landscapes (bottom-right panel in Figure \ref{fig:event_main}B). These results indicate that an external channel cannot solely explain the main findings.

\section{Discussion\label{sec:discussion}}

Climate change has had far-reaching impacts on human societies, shaping various aspects of human life such as agriculture, migration, and warfare (see \cite{ljungqvist2021} for a review). In contrast to previous studies, this study examines how climate has impacted the human mind and culture using unique data from paintings and artists. The first analysis revealed a significant association between the lightness of paintings and temperature anomalies. In addition, the second analysis uncovered the effects of temperature shocks on the brightness of paintings, with a more pronounced effect observed in paintings relying on artists’ imaginations rather than real objects. These findings indicate that climate has a non-negligible impact on the human mind and culture.

To understand the mechanisms behind significant social changes, such as political transitions and economic development, it is crucial to comprehend the influence of people’s minds and cultures that enable such transformations. While indicators of economic development, such as The Maddison Project \cite{bolt2020}, and indicators of political regimes, such as The Polity Project \cite{polity}, have been widely used in research, there has been no standard measure of people’s minds and cultures over an extended period. Although text data have been used in previous studies \cite{acerbi2013, morin2017, scheffer2021, iliev2016, kesebir2012, greenfield2013, grossmann2014, garg2018, hills2019, martins2020, jackson2019, choi2022}, there remain challenges in terms of coverage (both time and space) and content. This study demonstrated that art can serve as a valuable tool for analyzing people’s minds and cultures, complementing the existing data while providing unique insights.

The study has several limitations. First, older paintings are more likely to be available in Western countries. However, the spatial coverage becomes wider if paintings from more recent years are considered (see Figure \ref{fig:map_coverage} in the Supplementary Material). Future studies can also enrich the analysis by incorporating different art forms, such as sculptures and music. Second, while this study focuses on the lightness of paintings, there is scope for exploring other features of paintings in future research. Finally, while this analysis utilizes past data, it prompts an intriguing question about potential future trends for paintings. This aspect remains a topic for future research. \\

\noindent {\bf Acknowledgments.} This research was financially supported by the JSPS (18K12768, 21K13284). We want to thank Shuji Asaka, Hoshimi Hoashi, Shibao Mayo, Aiko Takehana, and Elizaveta Kugaevskaya for their valuable assistance with this research. In addition, we thank Takuma Kamada, Tsukasa Kodera, and Nils-Petter Lagerlöf for their valuable and insightful discussions and the seminar participants at Osaka and York for their helpful feedback. \\

\noindent {\bf Competing Interest Statement.} The author declares no conflict of interest.

\bibliographystyle{unsrtnat}
\bibliography{art}


\section*{Tables and Figures}


\begin{figure}[H]
\begin{center}
\includegraphics[width=0.95\linewidth]{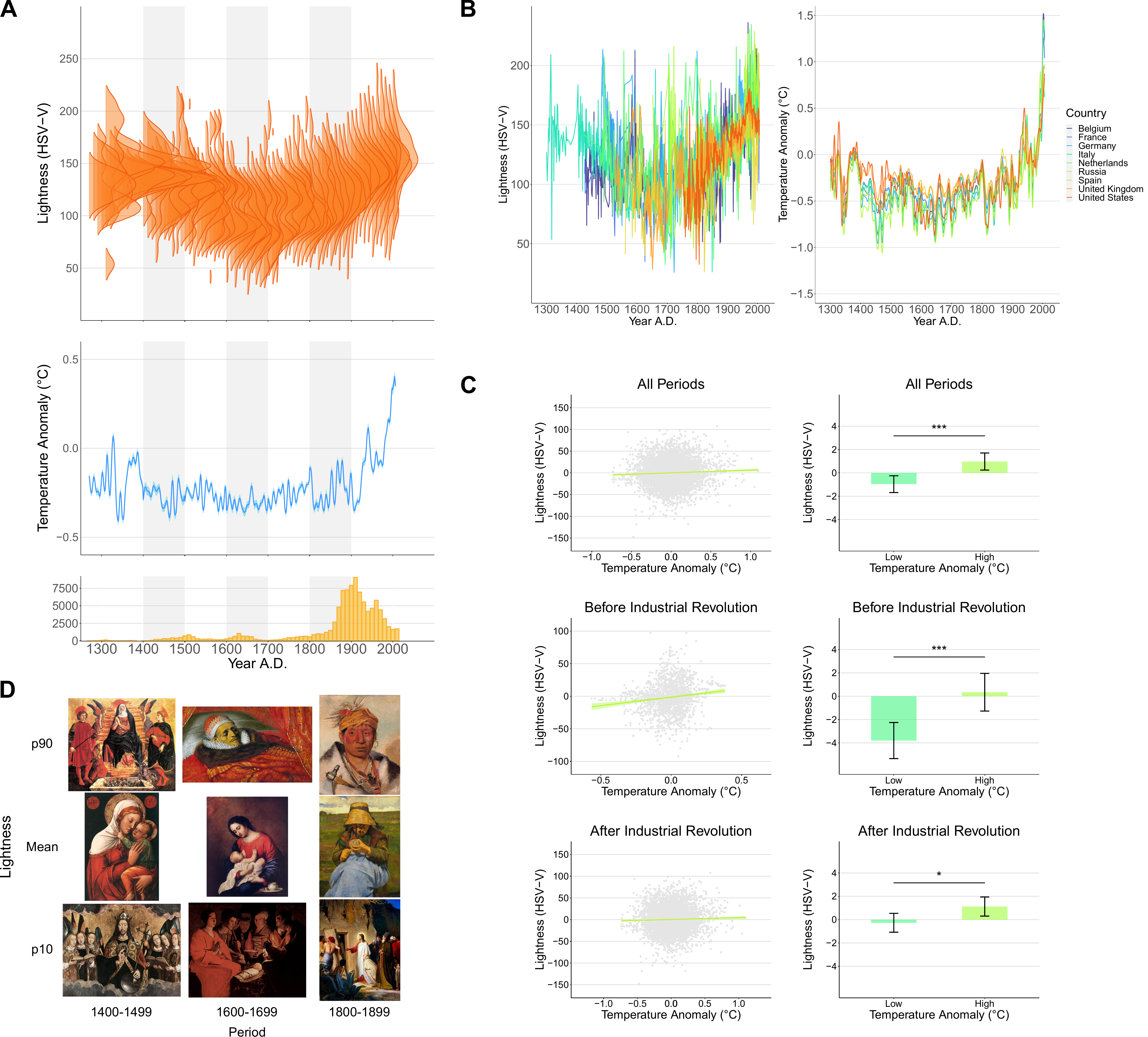}
\caption{Correlations between the lightness of paintings and temperature anomalies. (A) The top panel displays the density of lightness over a decade using country-level data. The middle panel shows the global average temperature anomalies and 95\% confidence intervals (CIs) by year using country-level data. The bottom panel indicates the number of paintings produced in each decade. The sample period is between 1270-2006. (B) Time series of painting lightness and temperature anomalies for selected countries (Belgium, France, Germany, Italy, Netherlands, Russia, Spain, United Kingdom, and the United States). (C) Left: Partial correlation between temperature anomalies and lightness after controlling for GDP per capita, population density, the presence of battles, country fixed effects, and art movement dummies. Each grey dot represents the average value for each country and year, and each green line indicates a fitting line with 95\% CIs. The top figure uses the full sample, the middle figure restricts the sample to pre-1750 periods, and the bottom figure restricts the sample to post-1750 periods. Right: The same as the respective left plots but categorize (residualized) temperature anomalies (x-axis) into high and low groups based on the median value. *$p$ $<$ 0.05, ** $<$ 0.01, and *** $<$ 0.001. The error bars indicate 95\% CIs. (D) Examples of paintings from the respective centuries colored grey in (A). The paintings at the top were brighter (90th percentile), those in the middle had average lightness, and those at the bottom were darker (10th percentile).\label{fig:main}}
\end{center}
\end{figure}

\clearpage

\clearpage

\clearpage

\appendix

\renewcommand{\thesection}{A.\arabic{section}} \setcounter{section}{0} %
\renewcommand{\thefigure}{A.\arabic{figure}} \setcounter{figure}{0} %
\renewcommand{\thetable}{A.\arabic{table}} \setcounter{table}{0}

\begin{mytitlepage}
\title{Supplementary Information for {\it Quantifying the Influence of Climate on Human Mind and Culture: Evidence from Visual Art}}\author{Shuhei Kitamura\thanks{%
ISER, Osaka University. Email: kitamura@iser.osaka-u.ac.jp.}}
\date{}
\maketitle
\end{mytitlepage}

\section{Materials and Methods\label{sec:methods}}

\subsection{Data sources and construction procedures}

The images of the paintings were obtained from Wikiart \cite{wikiart}. A total of 153,178 images and 2,836 artists were downloaded initially. Then, artworks from the Proto-Renaissance to modern art were retained, and the remainder were dropped. The following art genres were excluded: artists’ books, architecture, calligraphy, design, graffiti, installation, jewelry, mobile, mosaic, mural, ornament, performance, photo, quadratura, sculpture, sketch and study, stabile, tapestry, and video. Artists whose artworks were unavailable owing to copyright infringement issues were removed. This process yielded 126,310 images and 2,227 artists. Of these, approximately 21\% (26,351) of the artworks with missing information on the year of creation were removed. The final sample consisted of 99,959 images and 2,174 artists. For the remaining images with a range of years (e.g., 1752--1754) provided (7,366 images), the mid-year period (e.g., 1753) was used as the year of creation. The location of creation for most artworks lacking information was estimated based on the artist’s residential country at each point in time using the migration history of artists.

\begin{figure}[p]
\begin{center}
\includegraphics[width=\linewidth]{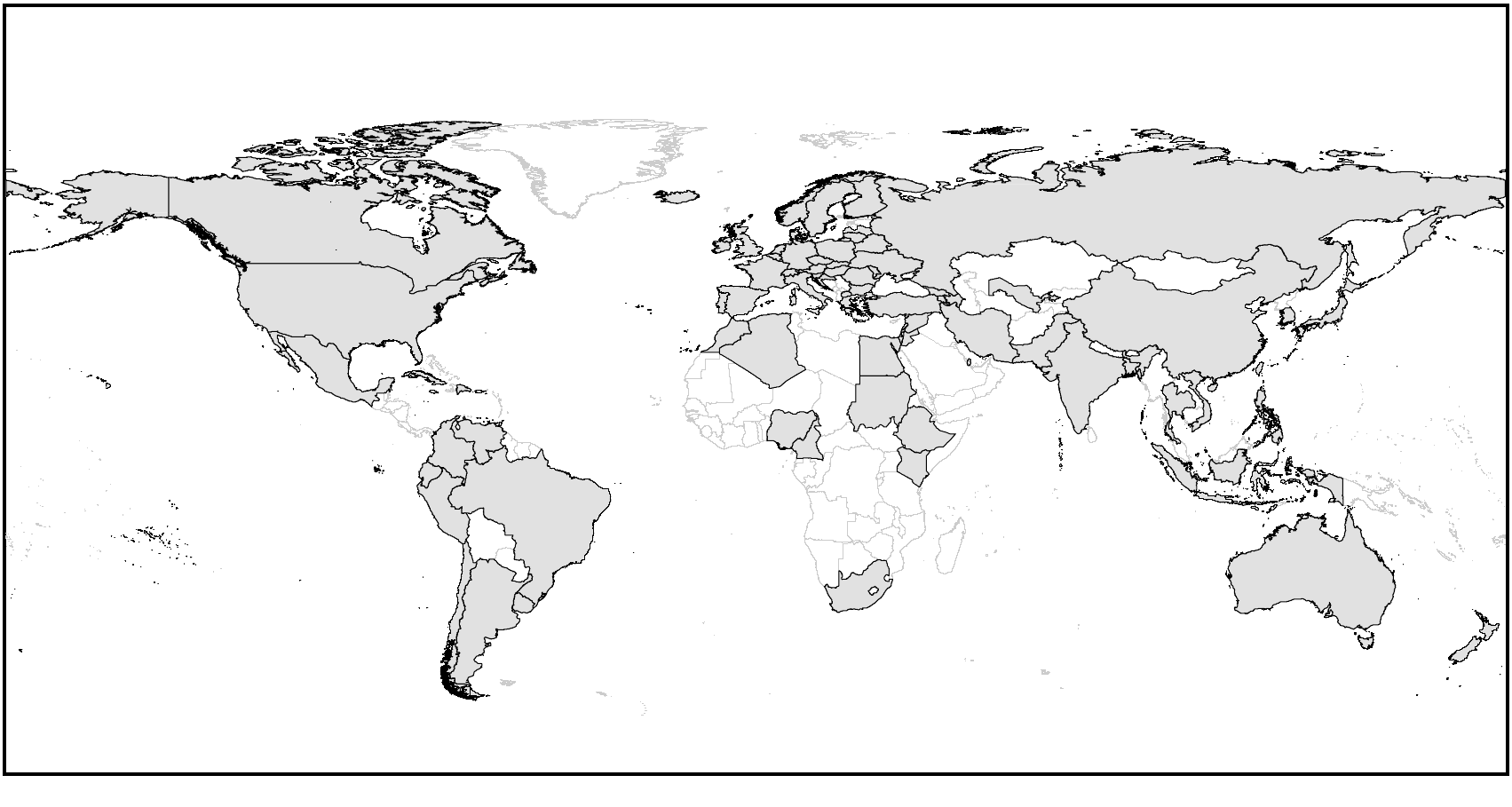}
\caption{Spatial coverage of the data. Shaded areas indicate the countries included in the data.\label{fig:map_coverage}}
\end{center}
\end{figure}

The biographies of artists, including their migration histories, were obtained from online sources, such as artists’ websites and online encyclopedias. In this study, migration is defined as the change in residential places between countries, but not within a country, as migration history is only available at the country level in many cases. Short trips between countries were not considered migration owing to the limited availability of such details. The country boundaries were based on GADM Version 3.6 \cite{gadm}. Figure \ref{fig:map_coverage} shows the spatial coverage of the data.

The lightness of the images was computed using Python OpenCV. The main analysis used the Hue-Saturation-Value (HSV) color space based on the Munsell color system. After importing the images into the BGR, they were converted to HSV. However, 12 images collapsed during this process, resulting in missing values. Therefore, the color information was obtained from 99,947 images. Missing observations in color information for the panel of artists were filled in using linear interpolation. 

For the main analysis, removing any range of pixels from images was avoided to prevent arbitrary truncation of certain brightness values. The results remain very similar even after removing very bright pixels (bottom-left figure in Figure \ref{fig:event_study_robust}).

Art movement dummies were created using the same dataset of paintings, which also contains information on each artist's association with specific art movements. However, an art movement could start too early if artists began the movement in the middle of their lifetimes but not from the beginning, or it could end too late if a recent individual artist revives a past movement. To address these issues, an art movement was defined as starting when there were two or more artists (regardless of where they live) in the same movement in the same year and ending when there were no longer two or more artists in the same movement in the same year. Then, the start and end years of an art movement were assigned to each country if at least one artist associated with that art movement defined above existed in that country. The years between the start and end were interpolated to fill in missing observations.

Other data sources and construction procedures were as follows: Global temperature data were taken from Mann et al. (2009) \cite{mann2009}, and values were resampled for smaller cells using the natural neighbor interpolation method. The average annual temperature was then calculated for each polygon. The geocoded battle data were taken from Kitamura (2022) \cite{kitamura2022}, matched with country polygons, and the number of battles per annum was counted within each country polygon. The gridded population density was obtained from Goldewijk et al. (2017) \cite{goldewijk2017}. The values were resampled for smaller cells using the natural neighbor interpolation method, with average values computed for each country polygon. Annual data were linearly interpolated for earlier periods where data were unavailable. Data on GDP per capita were taken from Bolt and van Zanden (2020) \cite{bolt2020} and linearly interpolated for missing observations. These country-level data were matched with each artist by year, based on their place of residence. All GIS processes were conducted using ArcMap Version 10.8. Table \ref{fig:summary_table} presents the summary statistics for the variables used in this study.

\subsection{Statistical Analysis}

\subsubsection{Correlation}

Pearson’s correlation coefficient and the {\it p}-value were reported for the correlational analysis between temperature anomalies and the lightness of paintings. The statistical analysis was performed using R 4.3.1. For all statistical tests, the significance level was set at 0.05.

\subsubsection{Event Study}
\begin{figure}
\begin{center}
\includegraphics[width=0.85\linewidth]{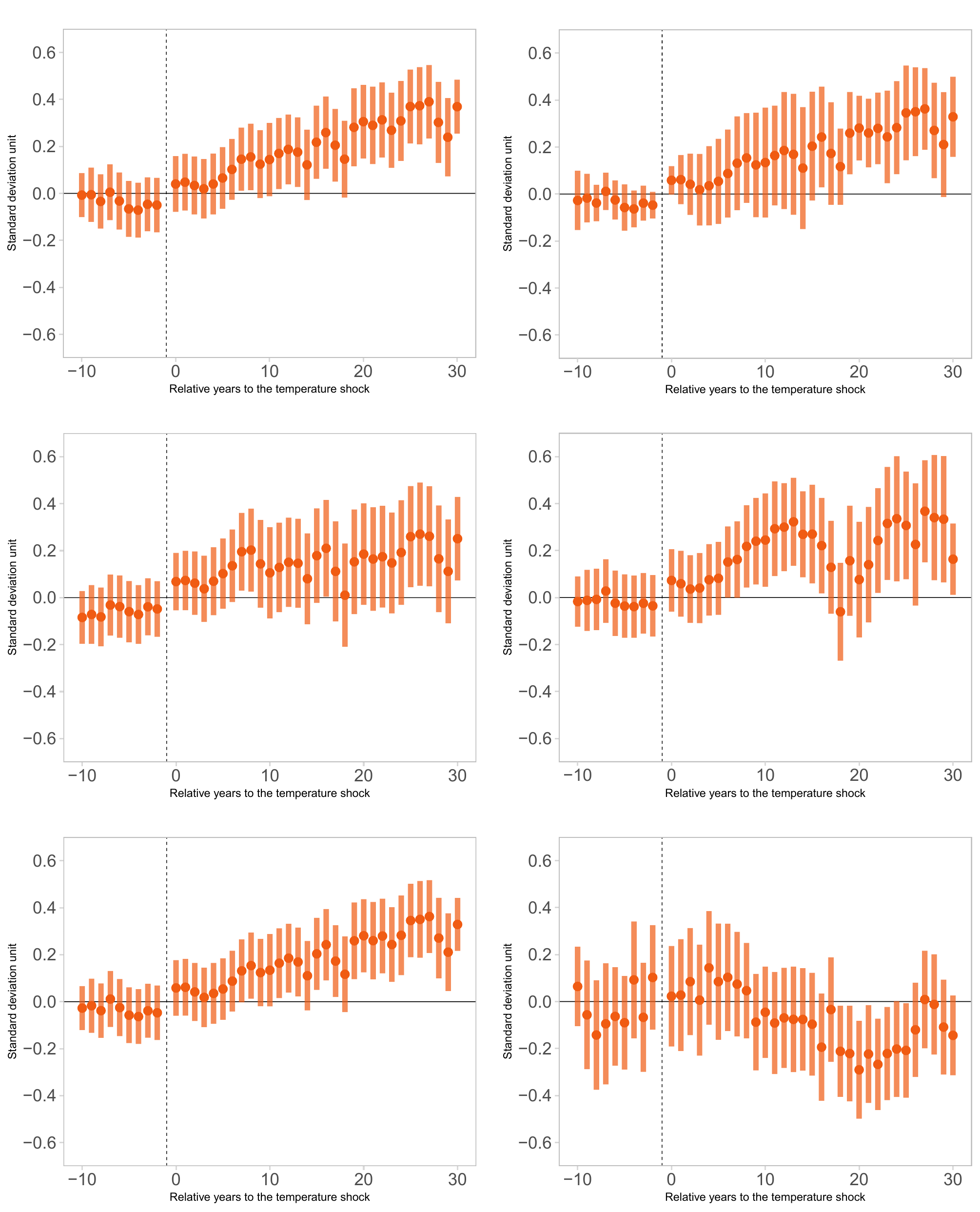}
\caption{Robustness checks. Top left: Estimated coefficients and associated 95\% CIs using equation (\ref{eq:event_main}) with additional control variables (GDP per capita, population density, and the presence of battles). Top right: Coefficients and associated 95\% CIs estimated using equation (\ref{eq:event_main}) with three-way clustered standard errors (artist, country, and year). Middle left: Estimated coefficients and associated 95\% CIs using the estimation method developed by Sun and Abraham (2021) \cite{sun2021}. Middle right: Estimated coefficients and associated 95\% CIs using equation (\ref{eq:event_main}) by excluding all artists whose treatment year was 1938. Bottom left: Estimated coefficients and associated 95\% CIs using equation (\ref{eq:event_main}) by excluding 250-255 tones on an 8-bit gray scale. Bottom right: Estimated coefficients and associated 95\% CIs using equation (\ref{eq:event_main}), except that the shocks are defined as cold-temperature shocks.\label{fig:event_study_robust}}
\end{center}
\end{figure}

The regression equation for the event study analysis is as follows:
\begin{equation}\label{eq:event_main}
Y_{art} = \sum_{l = -10, \neq -1}^{30} \beta_l I[t - k_{ar} = l] + \rho_a + \gamma_r + \tau_t + \epsilon_{art},
\end{equation}
where $Y_{art}$ is the dependent variable for artist $a$ living in country $r$ and year $t$; $I[t - k_{ar} = l]$ is an indicator that takes the value of 1 if the equation in brackets holds and 0 otherwise; $k_{ar}$ is the year that artist $a$ experienced a global hot temperature shock; $\rho_a$, $\gamma_r$, and $\tau_t$ are fixed effects; and $\epsilon_{art}$ is the error term. The top-left panel in Figure \ref{fig:event_study_robust} shows the results with additional controls: GDP per capita, population density, and the presence of battles in the country where artists were living. The main analysis uses robust standard errors, and the top-right panel in Figure \ref{fig:event_study_robust} shows the results using three-way clustered standard errors. The global hot-temperature shock was defined as the top 1\% of the temperature anomalies in the data. The first shock was administered to those who experienced two or more shocks. The distribution of these shocks is shown and discussed in Section \ref{sec:dist_shocks}.

The year before the shock was set as the baseline, and the coefficients $\beta_l$ capture the differences in outcome values between the treatment and control groups compared to the baseline year. To increase the interpretability of the results, the estimated coefficients were reported by dividing each coefficient by the standard deviation of the dependent variable. Finally, the original Wikiart data contained information on art genres, which was used to select paintings from particular art genres. All regression analyses were performed using STATA 15.

\subsubsection{DID estimation}
The regression equation for the analysis to examine the influence of personal attributes is as follows:
\begin{equation}\label{eq:event_hetero}
Y_{art} = \alpha_0 D_{art} + \alpha_1(D_{art} \times Z_{a}) + \phi_a + \psi_r + \xi_t + u_{art},
\end{equation}
where $Y_{art}$ is the dependent variable for artist $a$ living in country $r$ and year $t$, $D_{art}$ is an indicator that takes the value of one after the shock and zero otherwise, $Z_{a}$ is a personal attribute, $\phi_a$, $\psi_r$, and $\xi_t$ are fixed effects, and $u_{art}$ is the error term. The shock was defined the same way as in the main analysis. Personal attributes included age at the time of shock, sex (female = 1), source of income, and how the artist learned to paint. The attributes are selected following previous discussions with art historians but are limited by data availability. The underlying hypotheses were that the influence of hot temperatures might be greater for those exposed during their early childhood, for female artists, those with an unstable income, and those who had learned to paint by themselves. The coefficient $\alpha_1$ captures the heterogeneous responses to shocks based on each personal attribute (e.g., female artists’ responses to the shock compared to male artists’ responses to the same shock).

\section{Supplementary Text}

\subsection{Distribution of shocks}\label{sec:dist_shocks}

\begin{figure}
\begin{center}
\includegraphics[width=\linewidth]{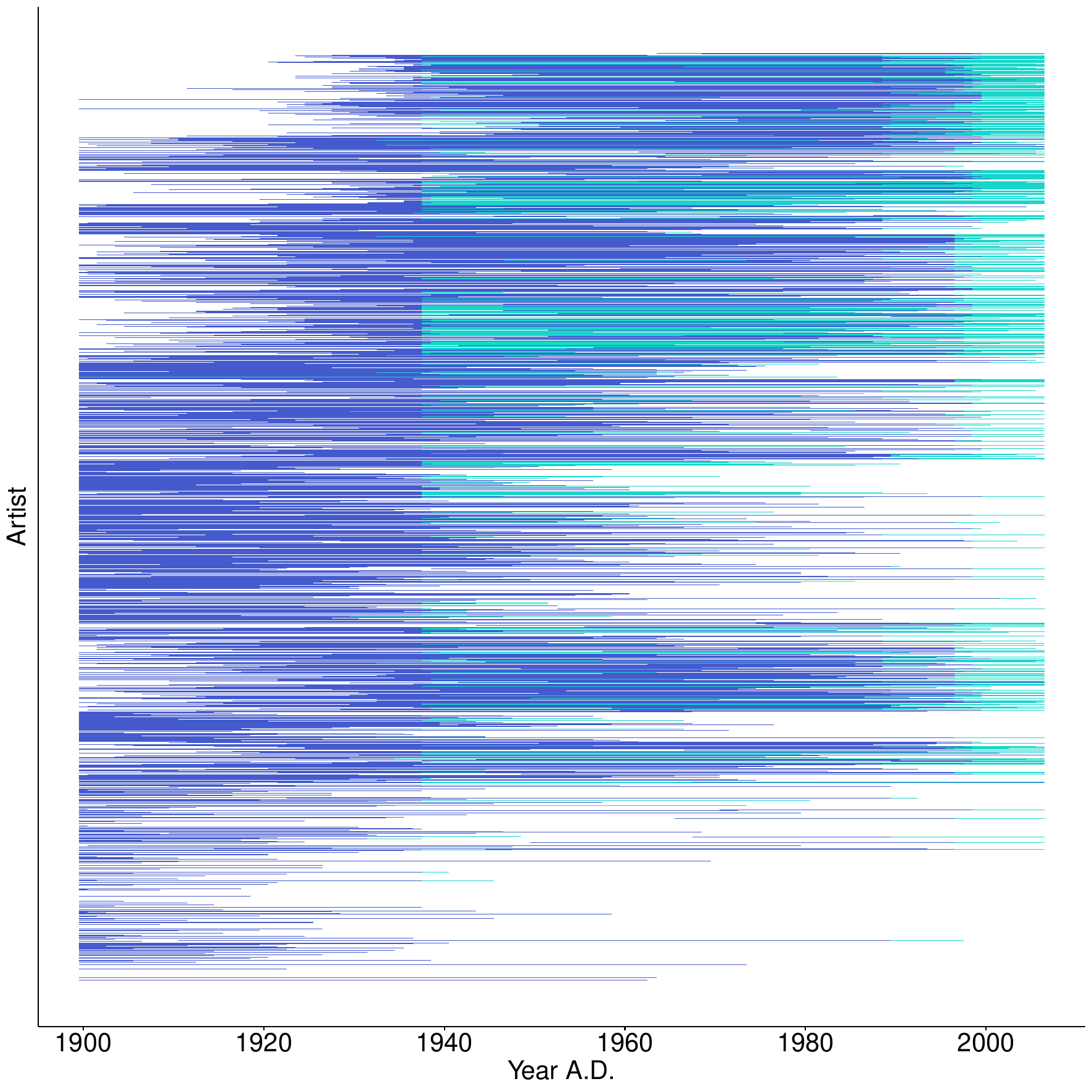}
\caption{Treatment and control status by artist, 1900-2006. A brighter blue indicates that artists have been treated, whereas a darker blue indicates that artists have not been treated (yet).\label{fig:dist_shocks}}
\end{center}
\end{figure}

Figure \ref{fig:dist_shocks} illustrates the artists’ treatment and control status, where a brighter blue indicates that artists have been treated, whereas a darker blue indicates that artists have not been treated (yet). As global temperatures increase, temperature shocks tend to occur in the 20th and 21st centuries (between 1900-2005). This figure also shows that a relatively substantial number of artists were exposed to the 1938 shock. To ensure this shock did not drive the main results, I ran a regression using equation (\ref{eq:event_main}) by excluding all artists whose treatment year was 1938. The middle-right panel of Figure \ref{fig:event_study_robust} shows that excluding relatively large samples does not systematically invalidate the main results. Thus, the results are unlikely to have been influenced by this particular shock.

\subsection{Effects of cold temperatures}\label{sec:cold}

In contrast to examining the effect of hot temperatures, examining the effect of cold temperatures is difficult owing to data availability. As the bottom panel of Figure \ref{fig:main}A in the main text indicates, more paintings have become available in recent years. Therefore, if the cold temperature events are defined first and then matched with artist data, only a few events (actually only two years) could be successfully matched with the artists’ data. This makes it almost impossible to conduct a rigorous econometric analysis.

One possible way to avoid this issue is to redefine the events of cold temperatures {\it after} merging the painting data. Therefore, the interpretation of the following regression estimates is the impact of cold temperature shocks {\it in the final sample}. With this caveat in mind, the bottom-right panel in Figure \ref{fig:event_study_robust} displays the coefficients from the regression, where zero indicates the year in which an artist was exposed to global cold temperatures. A global cold-temperature shock was defined as the bottom 1\% of temperature anomalies in the final sample.

Like hot temperature shocks, it takes several years for the effect to be statistically significant. The overall pattern was that artists tended to draw darker paintings after cold-temperature shocks. The magnitude of the change was 0.29 standard deviations ($p$ = 0.006) after 20 years following the shock. In contrast to the monotonic increase in the estimates in the case of hot-temperature shocks, the effects of cold-temperature shocks tend to diminish after reaching a peak. An interpretation of this result is that temperature shocks are not symmetrical; the effects of hot temperature shocks last relatively long, whereas cold temperature shocks have relatively short-lived effects.

\subsection{Effects on productivity}\label{sec:productivity}

\begin{figure}
\begin{center}
\includegraphics[width=0.8\linewidth]{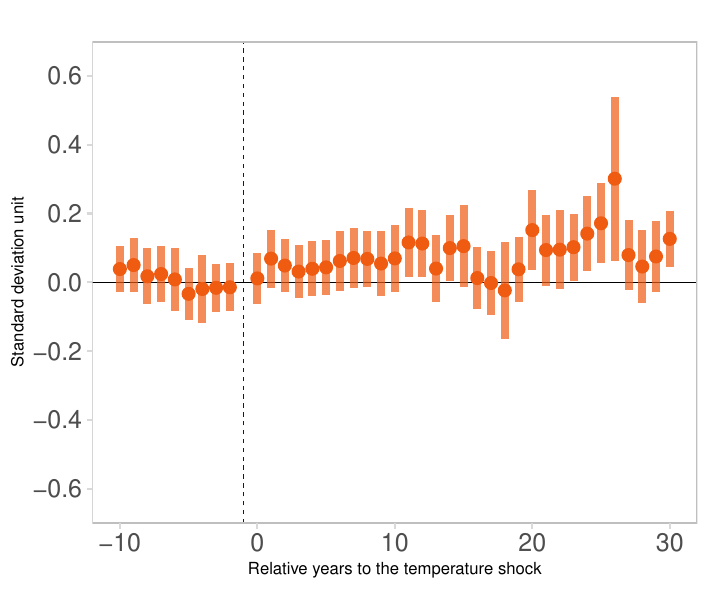}
\caption{Effects on productivity. The estimated coefficients and associated 95\% CIs using equation (\ref{eq:event_main}), with the number of paintings for each artist and year as the dependent variable.\label{fig:count_art}}
\end{center}
\end{figure}

The temperature may also affect the number of paintings an artist produces annually. To check this possibility, Figure \ref{fig:count_art} illustrates the estimates of regressing the equation (\ref{eq:event_main}) in the main text by replacing the dependent variable with a measure of ``productivity," that is, the total number of paintings for each artist and year. Missing observations in the number of paintings for the panel of artists were filled in using linear interpolation. The figure shows no clear evidence that temperature shocks affect artists’ productivity.


\clearpage 

\section{Supplementary Tables and Figures}

\vspace{6cm}

\begin{table}[!htbp] \centering 
\begin{tabular}{@{\extracolsep{5pt}}lccccc} 
\\[-1.8ex]\hline 
\hline \\[-1.8ex] 
 & \multicolumn{1}{c}{N} & \multicolumn{1}{c}{Mean} & \multicolumn{1}{c}{Std. Dev.} & \multicolumn{1}{c}{Min} & \multicolumn{1}{c}{Max} \\ 
\hline \\[-1.8ex] 
Age of the first shock & 50,979 & 50.74 & 24.25 & 0 & 96 \\ 
Female & 141,762 & 0.11 & 0.31 & 0 & 1 \\ 
Income source: Remittance & 141,837 & 0.03 & 0.17 & 0 & 1 \\ 
Income source: Patron & 141,837 & 0.10 & 0.31 & 0 & 1 \\ 
Income source: Employed & 141,837 & 0.55 & 0.50 & 0 & 1 \\ 
Art education: Self & 136,670 & 0.07 & 0.25 & 0 & 1 \\ 
Art education: Apprentice & 136,670 & 0.44 & 0.50 & 0 & 1 \\ 
Art education: Academy & 136,670 & 0.74 & 0.44 & 0 & 1 \\ 
Number of artworks & 66,634 & 2.24 & 3.75 & 1.00 & 191.00 \\ 
HSV-V & 66,634 & 137.13 & 39.14 & 5.96 & 254.32 \\ 
HSV-S & 66,417 & 91.29 & 36.56 & 0.0003 & 254.16 \\ 
HSV-H & 66,417 & 49.16 & 22.59 & 0.01 & 176.34 \\ 
HLS-L & 66,634 & 115.03 & 37.20 & 4.00 & 248.18 \\ 
CIELUV-L & 66,634 & 123.50 & 38.65 & 3.04 & 247.80 \\ 
Average global temperature anomalies & 141,837 & $-$0.15 & 0.31 & $-$1.41 & 1.52 \\ 
GDP per capita & 129,122 & 8,999.82 & 9,102.32 & 295.00 & 51,862.96 \\ 
Battles & 141,802 & 0.53 & 4.86 & 0 & 140 \\ 
Average population density & 141,837 & 72.07 & 67.76 & 0.01 & 2,247.82 \\ 
\hline \\[-1.8ex] 
\end{tabular} 
\caption{Summary statistics.\label{fig:summary_table}}
\end{table} 

\clearpage

\clearpage

\begin{figure}[p]
\begin{center}
\includegraphics[width=0.5\linewidth]{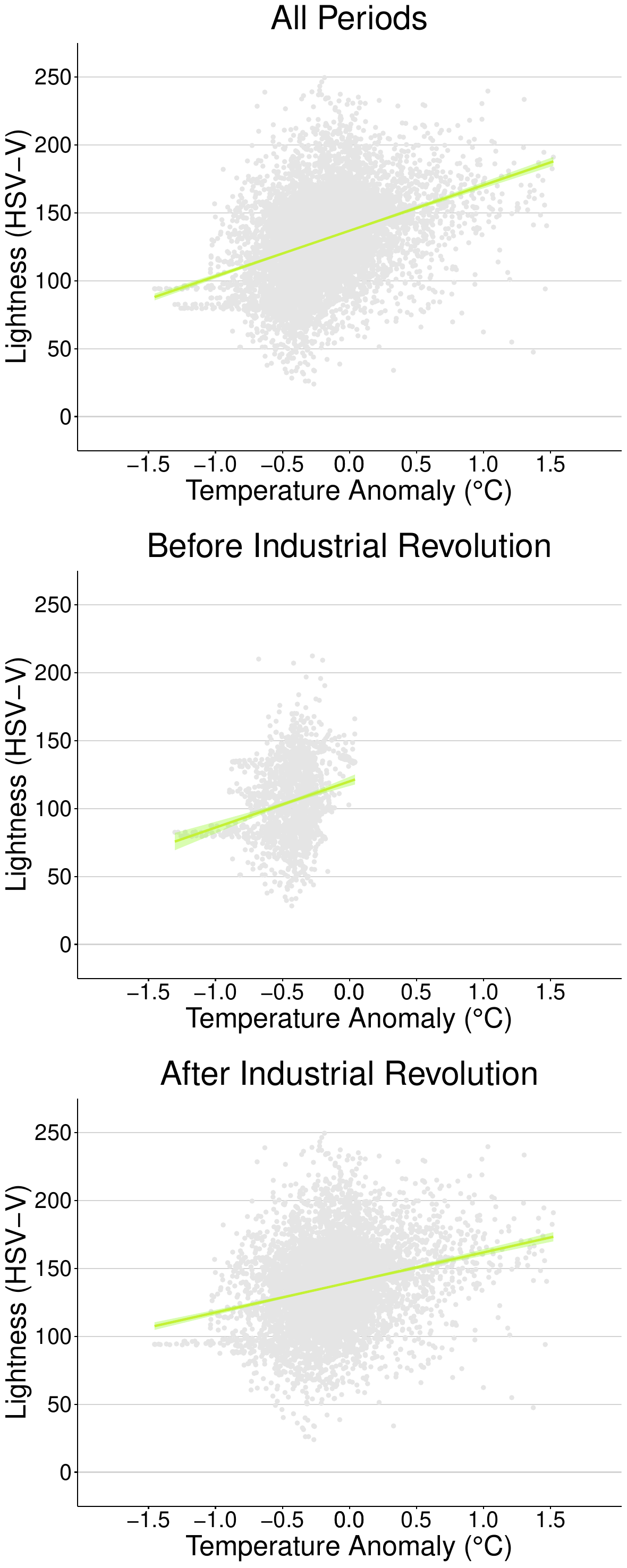}
\caption{Correlations between the lightness of paintings and temperature anomalies (without control variables). Each grey dot represents the average value for each country and year, and each green line indicates a fitting line with 95\% CIs. The top figure uses the full sample ($r$ = 0.355, $p$ $<$0.001), the middle figure restricts the sample to pre-1750 periods ($r$ = 0.221, $p$ $<$ 0.001), and the bottom figure restricts the sample to post-1750 periods ($r$ = 0.251, $p$ $<$ 0.001).\label{fig:corr_no_control}}
\end{center}
\end{figure}

\clearpage

\begin{figure}[p]
\begin{center}
\includegraphics[width=\linewidth]{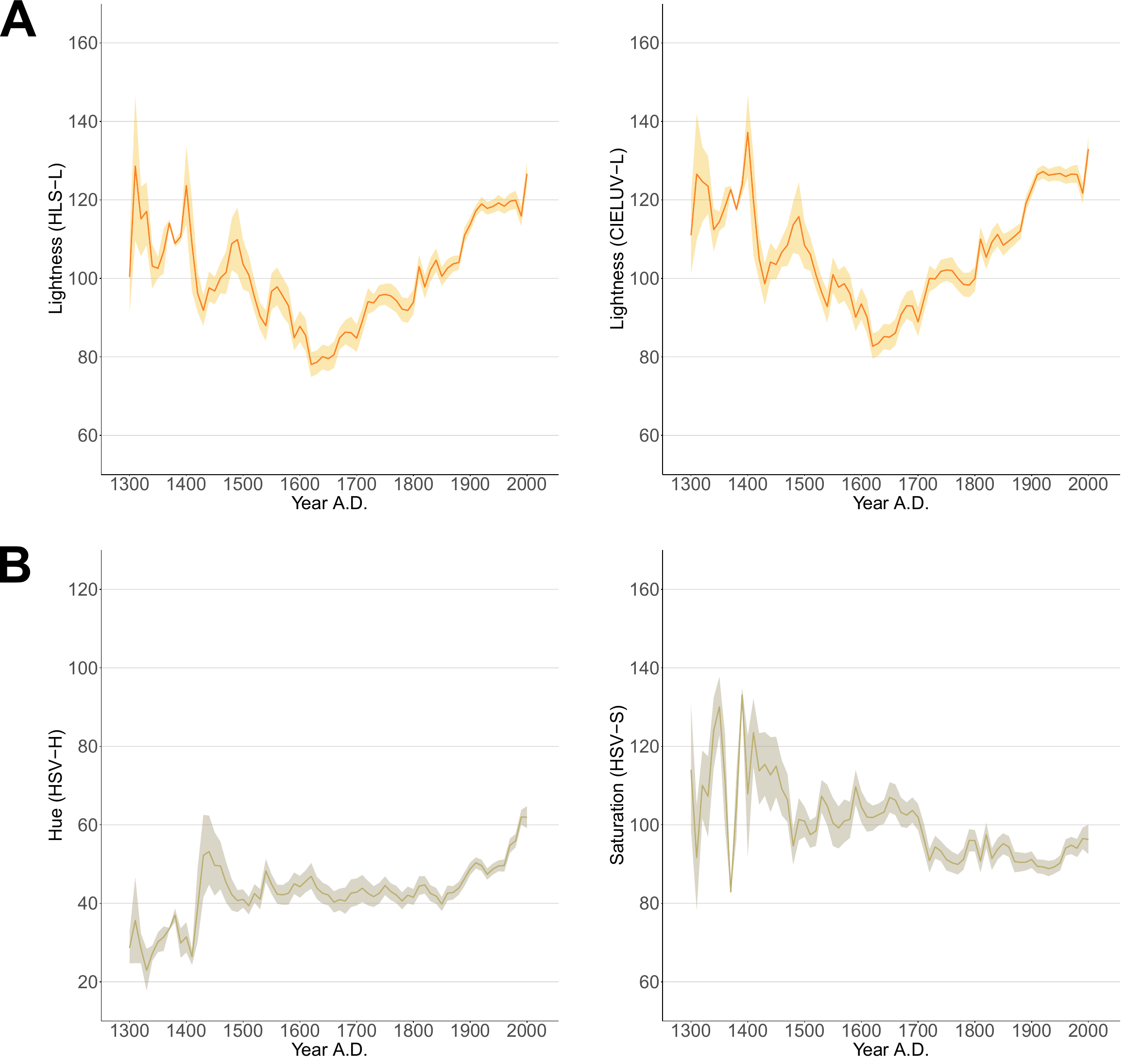}
\caption{Lightness of paintings using different color spaces and the hues and saturations of the paintings. (A) Left: The average lightness of paintings based on the HLS color space and 95\% CIs by year using country-level data. Right: Average lightness based on the CIELUV color space of paintings and 95\% CIs by year using country-level data. (B) Left: Average hues of paintings based on the HSV color space and 95\% CIs by year using country-level data. Right: Average saturation of paintings based on the HSV color space and 95\% CIs by year using country-level data.\label{fig:lightness_alt}}
\end{center}
\end{figure}

\clearpage

\clearpage

\clearpage

\end{document}